\documentclass[10pt,aps,pra,twocolumn,superscriptaddress,floatfix,nofootinbib,amsmath,amssymb]{revtex4-2}

\usepackage{graphicx}
\usepackage[normalem]{ulem}
\usepackage{bm}
\usepackage{braket}

\usepackage{xcolor}
\usepackage[
  colorlinks=true,
  linkcolor=blue,
  citecolor=blue,
  urlcolor=blue,
  breaklinks=true
]{hyperref}
\usepackage[capitalise]{cleveref}

\allowdisplaybreaks

\begin{document}
\title{Linearization Scheme of Shallow Water Equations for Quantum Algorithms}

\author{Till~Appel}
\thanks{These authors contributed equally to this work.}
\affiliation{Department of Information Technology and Electrical Engineering, ETH Zürich, Zürich, Switzerland}
\author{Zofia~Binczyk}
\thanks{These authors contributed equally to this work.}
\affiliation{Department of Information Technology and Electrical Engineering, ETH Zürich, Zürich, Switzerland}
\author{Francesco~Conoscenti}
\thanks{These authors contributed equally to this work.}
\affiliation{Department of Information Technology and Electrical Engineering, ETH Zürich, Zürich, Switzerland}
\author{Petr~Ivashkov}
\thanks{These authors contributed equally to this work.}
\affiliation{Department of Information Technology and Electrical Engineering, ETH Zürich, Zürich, Switzerland}
\author{Seyed~Ali~Hosseini}
\affiliation{Department of Mechanical and Process Engineering, ETH Zürich, Zürich, Switzerland}
\author{Ricardo~Garcia}
\email{quantumcomputing@moodys.com}
\affiliation{Moody’s, New York, NY, USA}
\author{Carmen~Recio}
\email{quantumcomputing@moodys.com}
\affiliation{Moody’s, New York, NY, USA}

\date{\today}

\begin{abstract}
Computational fluid dynamics lies at the heart of many issues in science and engineering, but solving the associated partial differential equations remains computationally demanding. With the rise of quantum computing, new approaches have emerged to address these challenges. In this work, we investigate the potential of quantum algorithms for solving the shallow water equations, which are, for example, used to model tsunami dynamics. By extending a linearization scheme previously developed in [\textit{Phys. Rev. Research} \textbf{7}, 013036 (2025)] for the Navier-Stokes equations, we create a mapping from the nonlinear shallow water equation to a linear system of equations, which, in principle, can be solved exponentially faster on a quantum device than on a classical computer. 
To validate our approach, we compare its results to an analytical solution and benchmark its dependence on key parameters. Additionally, we implement a quantum linear system solver based on quantum singular value transformation and study its performance in connection to our mapping. Our results demonstrate the potential of applying quantum algorithms to fluid dynamics problems and highlight necessary considerations for future developments. 
\end{abstract}

\maketitle

\section{Introduction}\label{sec:introduction}
Computational fluid dynamics (CFD) underpins a wide range of applications in engineering and science, from aerodynamic design to weather forecasting. However, solving the governing partial differential equations (PDEs) remains computationally demanding, even with the capabilities of today’s most advanced supercomputers. 

Quantum computing potentially offers asymptotic advantages for certain structured linear algebra tasks due to its ability to manipulate vectors in exponentially large Hilbert spaces in polynomial time~\cite{nielsen2010quantum}. A famous example is the Harrow–Hassidim–Lloyd (HHL) algorithm that achieves an exponential speedup in solving linear systems of equations~\cite{harrow2009quantum}. Subsequent works extended this quantum advantage to linear differential equations~\cite{berry2014high, berry2017quantum}. One may ask whether such methods could be used to solve fluid dynamics where the underlying PDEs are reduced to large linear systems~\cite{succi2023quantum}. It should be immediately noted that claims of exponential speedup should be interpreted cautiously, as the cost of loading the classical data into quantum states and later retrieving the result through measurement can offset the claimed advantage~\cite{aaronson2015read}. Much of the research at the intersection of CFD and quantum computing is centered on the Navier-Stokes equations (NSE), the fundamental equations governing fluid motion, as, for example \cite{bharadwaj2024simulating, yepez1998quantum, gaitan2020finding, gaitan2021finding, PhysRevResearch.5.033182, fluids1020018, syamlal2024computational}. Approaches based on the Lattice Boltzmann method, a promising and increasingly popular method for quantum CFD, were focused on in e.g. \cite{li2025potential, sanavio2024lattice, itani2023quantum, itani2022analysis, bakker2023quantum}. Several approaches based on these equations show potential, with some promising exponential speedups given specific assumptions, such as moderate Reynolds numbers \cite{sanavio2024lattice} or periodic boundary conditions with weakly compressible fluid (i.e. low Mach number) and decaying turbulence \cite{li2025potential}. These methods often rely on tools and techniques tailored to the NSE, limiting their applicability to other PDEs. The question of whether certain mappings are applicable to a broader class of CFD problems remains largely unanswered. 

In this work, we extend the application of quantum computing in CFD by adapting a linearization scheme, that was originally developed by Li et al. \cite{li2025potential} for the NSE, to the shallow water equations (SWE). The SWEs play a central role in modeling free-surface flows in oceans, rivers, and the atmosphere. Unlike the full three-dimensional NSE, the SWEs provide an efficient, depth-averaged model while still capturing essential nonlinear wave phenomena. Nonetheless, solving SWEs at high resolution or over long timescales remains computationally intensive, motivating the exploration of potential speedup though quantum computing.

A significant challenge in simulating PDEs such as the SWE on quantum hardware lies in addressing the nonlinearities as quantum operations are inherently linear. To overcome this, we transform the SWE into a linear system of equations (LSE) through a series of steps adapted from \cite{li2025potential}.

First, in \cref{sec:lbm}, the non-linearities are made local by using discrete velocity Boltzmann equations recovering the SWE in the hydrodynamic limit. Using the Carleman linearization, the local non-linearity is replaced by additional degrees of freedom (\cref{sec:clin}).
This transformation converts the problem into a high-dimensional system of ordinary differential equations (ODEs) and leverages the fact that degrees of freedom scale exponentially with the number of qubits in quantum computers. Next, the forward Euler method is applied in \cref{sec:forward_euler} to generate an LSE suitable for mapping onto quantum linear system solvers (QLSS), such as the quantum singular value transform (QSVT). These algorithms efficiently perform matrix inversion to solve equations of the form $Ax = b$, where the matrix $A$ and vector $b$ (representing the initial state in our case) are known.

Apart from the mathematical adaptation of the mapping from Li. et al to the SWE and their additional non-linearity, we performed the first extensive benchmarking scheme for the validity of this kind of linearization scheme in \cref{sec:classical_benchmarking}. Our results demonstrate that the approach accurately reproduces the analytically predicted dynamics. Since no sufficiently large quantum platform is currently available, these tests were carried out on a classical computing cluster. However, because the mapping is designed for the high-dimensional Hilbert spaces offered by quantum architectures, the range of test cases we can explore by simulating a quantum computer on a classical machine is inherently limited. Despite this constraint, our findings indicate the potential for efficient linearization schemes and robust mappings to quantum hardware with promising applications beyond specific PDEs like the NSE and SWE.

In addition to the classical simulations, we also implement a proof-of-principle QSVT algorithm in \cref{sec:qlss} to explore the relationship between the mathematical transformations and the requirements of quantum circuits. Specifically, we analyze how the condition number—the key parameter governing the computational cost of quantum linear system solvers—scales with the parameters of the modeled system.

\section{Shallow water equations}\label{sec:swe}
Fundamentally, the SWE are a coupled set of hyperbolic partial differential equations derived by vertically integrating the incompressible Navier-Stokes equations~\cite{toro2024computational,SWE-book}. They are widely used in fluid dynamics to describe fluids or gases in hydrostatic balance whose vertical dynamics are negligible compared to horizontal behavior. This assumption typically applies to systems with shallow depths but can also extend to deeper environments where surface-level dynamics dominate, such as in tsunami propagation on open water \cite{SWE-book}.

In this proof-of-principle implementation, we target the one-dimensional SWE in their simplest form, as represented by \cref{SWE-1} and \cref{SWE-2}:
\begin{align}
\label{SWE-1}
\frac{\partial h}{\partial t} + \frac{\partial (hu)}{\partial x} = 0
\end{align}

\begin{align}
\label{SWE-2}
\frac{\partial (hu)}{\partial t} + \frac{\partial \left(hu^2 + \frac{1}{2}gh^2\right)}{\partial x} = 0
\end{align}
In these equations, $h=h(x,t)$ denotes the water depth and $u=u(x,t)$ the horizontal velocity, both of which vary as functions of space ($x$) and time ($t$). The constant $g$ represents gravitational acceleration.
More elaborate formulations of the SWE include factors such as water surface elevation, frictional losses, or two spatial dimensions.
\section{Discrete velocity Boltzmann equations}\label{sec:lbm}
The discrete velocity Boltzmann system of equations (DVBE) is the basis for numerical approaches, such as the lattice Boltzmann method, to solve the Navier-Stokes equations. It consists of a system of non-homogeneous hyperbolic PDEs for a set of discrete probability distribution functions, $f_i$, along discrete particle speed vectors $c_i$,
\begin{equation}\label{eq:DVBE}
    \frac{ \partial f_i}{\partial t} + \bm{c}_i\cdot  \bm{\nabla} f_i = \frac{1}{\tau}\left(f_i^{\rm eq} - f_i\right), i=1,\dots,Q.
\end{equation}
where $\tau$ is the relaxation time and $f_i^{\rm eq}$ are the discrete equilibrium distribution functions. Note that the evolution of the system is driven by two contributions: (a) streaming operator (second term on the left hand side) and (b) collision operator (term on the right hand side). Here $Q$ is the number of discrete velocities. In the context of the present study,  as for isothermal Navier-Stokes equations, $Q=d^3$ where $d$ is the dimensionality of space. While the scheme, as derived from kinetic theory, was initially developed to recover the incompressible Navier-Stokes system in the hydrodynamic limit, proper tuning of the discrete equilibrium and definition of the relaxation time can allow for application to other macroscopic balance equations, such as the shallow water equations discussed above. In the context of the shallow water equations, water depth and corresponding momentum being the conserved variables, they are invariants of the collision operator and are computed as,
\begin{equation}\label{eq:mom_0}
    \sum_{i=1}^Q f_i = \sum_{i=1}^Q f_i^{\rm eq} = h,
\end{equation}
and,
\begin{equation}\label{eq:mom_1}
    \sum_{i=1}^Q \bm{c}_i f_i = \sum_{i=1}^Q \bm{c}_i f_i^{\rm eq} = h \bm{u}.
\end{equation}
The literature on the shallow water equations is especially rich in the context of the lattice Boltzmann method with the first successful models proposed as early as in the late 90's and early 2000's, see~\cite{salmon1999lattice,dellar2002nonhydrodynamic,zhou2002lattice}. The model proposed by Salmon, forms the basis for almost all models developed using the lattice Boltzmann method and relies on an equilibrium distribution function defined as --for a 2D, nine discrete velocity lattice,
\begin{multline}\label{eq:2nd_equilibrium}
    f_i^{\rm eq} = w_i h\left(1+\frac{\bm{c}_i\cdot\bm{u}}{\varsigma^2}+\frac{1}{2\varsigma^4}\left(\left(g h/2-\varsigma^2\right)\bm{I} +\bm{u}\otimes\bm{u}\right) \right.\\ \left. :\left(\bm{c}_i\otimes\bm{c}_i - \varsigma^2\bm{I}\right)\right) + w_i^* \left(\frac{1}{4}h - \frac{3}{8}g h^2\right),
\end{multline}
where, $w_i$ is a weight associated to each discrete velocity of index $i$. The vector of all weights is
\begin{equation}
    w = \left[ \frac{4}{9}, \frac{1}{9}, \frac{1}{9}, \frac{1}{9}, \frac{1}{9}, \frac{1}{36}, \frac{1}{36}, \frac{1}{36}, \frac{1}{36} \right],
\end{equation}
with a set of modified weights $w_i^*$ given by,
\begin{equation}
    w^* = \left[ \frac{4}{9}, -\frac{2}{9}, -\frac{2}{9}, -\frac{2}{9}, -\frac{2}{9}, \frac{1}{9}, \frac{1}{9}, \frac{1}{9}, \frac{1}{9} \right],
\end{equation}
Here, $\varsigma=1/\sqrt{3}$ is the lattice speed of sound. Salmon's equilibrium differs from a second order polynomial equilibrium only in the last term, which affects the contracted fourth-order equilibrium moment, and as discussed by Dellar in \cite{dellar2002nonhydrodynamic}, allows for more stable simulations in 2D. In the context of 1D systems of interest here, i.e. the D1Q3 lattice, both approaches reduce to the same form, i.e.,
\begin{equation}\label{eq:2nd_equilibrium_1d}
    f_i^{\rm eq} = w_i h\left(1+\frac{c_i u }{\varsigma^2} + \frac{\left(g h/2-\varsigma^2 + u^2\right)\left(c_i^2 - \varsigma^2\right)}{2\varsigma^4}\right).
\end{equation}
with weights
\begin{equation}
    w = \left[ \frac{2}{3}, \frac{1}{3}, \frac{1}{3} \right]
\end{equation}
as in \cite{VANTHANG20107373}. It can be shown, through a multi-scale analysis that this equilibrium along with $\tau=2\nu/g h$, where $\nu$ is the kinematic viscosity, would lead to the recovery of the proper Euler level shallow water equations, i.e. \cref{SWE-1} and \cref{SWE-2}. As such, the 1D discrete velocity Boltzmann system of equations relying on three discrete velocities $c_i\in\{-1,0,1\}$, \cref{eq:DVBE}, along with the 1D discrete equilibrium in \cref{eq:2nd_equilibrium_1d} and \cref{eq:mom_0} and \cref{eq:mom_1} will form the basis for our 1D SWE solver.

\subsection{Note on conservation terms}\label{sec:conservation_terms}
Note that the DVBE in \cref{eq:DVBE} is not the formula for the Lattice Boltzmann method strictly speaking as there are no conservation terms (see  \cite{hosseini2023lattice} for more details). As such, the approach used in this paper is a finite difference method which was chosen for simplicity for proof of concept and for wider applicability as an approximative model.
\section{Carleman linearization of LBE}\label{sec:clin}
The method outlined in the following section adopts the notation and is based on Ref.\ \cite{li2025potential} but is applied to the SWE rather than the NSE.

\subsection{For one grid point}\label{sec:one_grid_point}
The next step is to do the Carleman linearization. This step removes the nonlinearity from the collision operator by translating the nonlinearity from the operators to the variables, and is essentially a change of variable from the set of discrete probability distribution functions $\bm{f} = (..., f_i, ..)$ to the vector $V $ that contains the permutations of the components of $\bm{f}$. It is defined as:
\begin{equation}
\begin{split}
    V &= (f_1, f_2, f_3, f_1^2, f_1f_2, f_1f_3,f_2f_1...)^\mathsf{T}\\
    &= (\bm{f}, \bm{f}\otimes\bm{f}, \bm{f}\otimes\bm{f}\otimes\bm{f}, ...)^\mathsf{T}\in \mathbb{R}^{\infty}.
\end{split}
\end{equation}

It is worth highlighting that, as will become evident in the calculations below, the maximal possible order of permutation of $\bm{f}$ is the same order as the maximal value of the exponent on $f_i$ by the nature of how the Carleman linearization is defined. Due to the exact form of the equilibrium distribution functions \cref{eq:DVBE}, after injecting the statistical moments \cref{eq:mom_0} and \cref{eq:mom_1}, the exponents on $f_i$ are infinite which directly results in an infinite order of permutations and infinite dimension of $V$. 

The goal is, after some substitutions, to rearrange the DVBE to arrive at the following first-order ODE:
\begin{align}
    \frac{\partial V^{(k)}}{\partial t} =C^{(k)}V^{(k)},
\end{align}
where ${C}^{(k)}$ is the Carleman matrix defined as 
\begin{align}
    {C}^{(k)} = C_s^{(k)} + C_c^{(k)}
\end{align} 

where $C_s^{(k)}$ and $C_c^{(k)}$ are the streaming and collision matrices of truncation order $k$, respectively. The truncation order translates directly to the order of permutation, and polynomial order of $f_i$, that we decide to keep. The derivation of these matrices and a discussion of the truncation order are detailed in the following subsections.

The dimension of ${V}^{(k)}$ is $\sum_{j=1}^{k}Q^j$ where $Q$ is the number of discrete velocities or vector components of the lattice. ${C}^{(k)}$ is a square matrix of dimension $(\sum_{j=1}^{k}Q^j)^2$.

To start, the change of variable is achieved after injecting the statistical moments \cref{eq:mom_0} and \cref{eq:mom_1} into the explicit expressions of $f_i^{\rm eq}$ in \cref{eq:DVBE} for each $i=1,\dots,Q$ and rearranging the equations to create one equation as a function of matrices. For the following, to create the collision and streaming matrices, we consider the terms generated from the collision and streaming operators in \cref{eq:DVBE} separately. After injecting the statistical moments, we rearrange and collect powers of $f_i$ without combining terms from the collision and streaming operators. We consider the vector equation in $\bm{f}$ and build the matrices that multiply each power of $\bm{f}$ that will be labeled as $F^{(j)}$. The resulting equation is of the form:
\begin{align} \label{eq:matLBE}
    \frac{\partial \bm{f}}{\partial t} = \overbrace{-\bm{c}\cdot  \bm{\nabla} \bm{f}}^{streaming} + \overbrace{F^{(1)}\bm{f} + F^{(2)}\bm{f}^{[2]} + F^{(3)}\bm{f}^{[3]} + ...}^{collision}
\end{align}
where 
\begin{align}
    \bm{f}^{[j]} = \overbrace{\bm{f}\otimes...\otimes \bm{f}}^{j \textrm{ times}},
\end{align}
and $[j]$ indicates the degree of the Kronecker product and $\bm{c} = (c_1, ...,c_i,...)$ the vector of the discrete particle speed vectors.
\subsubsection{Truncation}
It is important to consider the order of truncation; otherwise, the dimension of the vector $V$ and the Carleman matrices is potentially infinite to accommodate all permutations of $\bm{f}$. The truncation order is, by construction, equal to the maximal order of permutation of $\bm{f}$, which comes directly from the maximal exponent on $f_i$ due to how this transformation is defined and as can be seen in the subsequent chapters. In practice, the truncation order can be chosen to limit computational complexity, and here we explain the consequences it has on the physical constraints of the model.

In this part of the work, some assumptions differ from the work in Ref.\ \cite{li2025potential}. We similarly choose a truncation of order 3, but here it is explicitly to limit the size of the Carleman matrices for computational feasibility (see before, ${C}^{(k)}$ is a square matrix of dimension $(\sum_{j=1}^{k}Q^j)^2$), while in \cite{li2025potential} it is a more natural choice based on physical assumptions of the NSE. 

In the case of the SWE, this choice introduces a key limitation: when truncating at order 3, only flows with small variations in height can be modeled accurately. This is due to the substitution of statistical moments for $h$ and $u$ into the equilibrium distribution functions. This is because of the term $hu^2$, it can be rewritten as follows: $hu^2= (hu)^2(1/h)$ which allows a direct substitution of the statistical moments. The term $1/h$ is computationally awkward and so can be replaced by a Taylor expansion to a potentially infinite order of precision, before finally injecting the statistical moments.

Recall that $h$ and $hu$ are both linear functions of $f_i$ in the statistical moments. Let the Taylor expansion of $1/h$ to be order $t$, thus giving $f_i^t$ as the highest order term. $(hu)^2$ results in a term in $f_i^2$, thus $hu^2= (hu)^2(1/h)$ effectively increases the order of the $f_i$ to $2+t$. This is the only source of high order terms in the system of equations, and thus, $2 +t$ is the maximal order of the system, the maximal exponent of $f_i$ and the maximal order of permutation of $\bm{f}$. Let the truncation order be labeled $k$, this gives $ 2 +t = k$, thus $ t= k-2$, so truncating at order $k$ effectively limits the order of the Taylor expansion of $1/h$ up to order $k - 2$. 

Specifically, with a truncation order of 3, only the linear (first-order) terms in the Taylor polynomial are retained, which is a good approximation of $1/h$ only for small variations of $h$. Consequently, this model is only accurate for flows where $h$ exhibits minimal deviation from a nominal value. This can be expressed as (for normalized $h$) :
\begin{align}
    \frac{1}{h} \approx 2- h, \ \textrm{for}\ |1-h|\ll1.
\end{align}
This restriction is not a significant problem since many interesting scenarios can still be modeled, such as the propagation of tsunami waves where the additional height of the wave is very small compared to the total depth of the ocean.

To be able to model systems where the height varies more significantly, it is necessary to choose a larger truncation order, thus allowing more terms of the Taylor expansion to remain in the final expression. 

It is interesting to note that in the case of using the DVBE to solve the NSE as in Ref.\ \cite{li2025potential}, the choice to use a truncation of order 3 appears more naturally. In the NSE the parameter $h$ is replaced by the fluid density $\rho$ (the term $1/\rho$ appears after injecting the statistical moments). In the case of incompressible fluids, it is always true that the density does not vary significantly from the initial density, and because of this, the Taylor expansion of $1/\rho$ truncated to first order is done at the beginning of the calculation for simplification. This causes terms in maximal order $\bm{f}^{[3]}$ to appear. This means that to accommodate the first order precision of the Taylor expansion, the truncation order must be 3. As incompressible flows are very often chosen to be modeled, this is not an uncommon or significantly limiting approximation.
\subsubsection{Error due to truncation order}
This subsection also deviates from Ref.\ \cite{li2025potential}.

In \cite{itani2022analysis} Itani and Succi prove that the error improves exponentially with the Carlemann truncation order. The calculation of error due to truncation order in terms of the parameters of the system in this paper can be done as explained in the following. 

In order to properly quantify the theoretical error, it is essential to first understand what terms exactly cause the higher order permutations of $\bm{f}$. This comes from the step where the statistical moments \cref{eq:mom_0} and \cref{eq:mom_1} are injected into the explicit expressions of $f_i^{\rm eq}$ in \cref{eq:DVBE}, and thus the error depends on the exact expressions of $f_i^{\rm eq}$ which are different for different lattice types. In this work, we choose to simulate a D1Q3 lattice so the error analysis will be done with this lattice in mind, but the method is analogous for all lattice types. 

The higher order permutations come from the term $hu^2$. As explained in the previous section, this term causes a Taylor polynomial of $1/h$ to an infinite degree to appear in order to allow the substitution of the statistical moments. Thus, truncation of the Carlemann matrices is reduced to truncation of this Taylor polynomial. As explained previously, the truncation order $k$ is 2 higher than the precision order of the Taylor expansion. Thus, for a truncation of order $k$, the error is $\mathcal{O}\left(\left| 1-h \right|^{k-2}\right) $ for normalized $h$.
\subsubsection{Collision matrix}{}
The collision matrix is derived by considering the matrices $F^{(j)}$ in the following terms from equation \cref{eq:matLBE}:
\begin{align}
     F^{(1)}\bm{f} + F^{(2)}\bm{f}^{[2]} + F^{(3)}\bm{f}^{[3]} + ...
\end{align}
By the Carleman method, the collision matrix is of the form:
\begin{equation} \label{eq:collmat}
C_c^{(k)} = \resizebox{0.35 \textwidth}{!}{
 $ \begin{bmatrix}{}
  A_1^1 & A_2^1 & A_3^1 & 0 & 0 & \ldots & 0 & 0 & 0 & 0 \\
  0 & A_2^2 & A_3^2 & A_4^2 & 0 & \ldots & 0 & 0 & 0 & 0 \\
  \vdots & \vdots & \vdots & \vdots & \vdots & \ddots & \vdots & \vdots & \vdots & \vdots \\
  0 & 0 & 0 & 0 & 0 & \ldots & 0 & A_{k-2}^{k-2} & A_{k-1}^{k-2} & A_k^{k-2} \\
  0 & 0 & 0 & 0 & 0 & \ldots & 0 & 0 & A_{k-1}^{k-1} & A_k^{k-1} \\
  0 & 0 & 0 & 0 & 0 & \ldots & 0 & 0 & 0 & A_k^k
  \end{bmatrix}$,
}
\end{equation}
with the transfer matrices defined as:
\begin{equation}
\label{eq:A}
A_{j+m-1}^i = \resizebox{0.3 \textwidth}{!}{
  $\sum_{r=1}^j \overbrace{\mathbb{I}_{Q\times Q} \otimes \cdots \otimes{\underset{\substack{\uparrow \\ \text { r-th position }}}{F^{(m)}} \otimes \cdots \otimes \mathbb{I}_{Q\times Q}}}^{i\text { factors }}$,
}
\end{equation}
where $\mathbb{I}_{Q\times Q}$ is the $Q \times Q$ identity matrix and $k$ is the chosen truncation order, in this case $k=3$. 
\subsubsection{Streaming matrix}
Considering \cref{eq:DVBE}, the streaming matrix is built by first defining a matrix $S$ such that:
\begin{equation} \label{eq:S}
    S\bm{f} = -\bm{c}\cdot  \bm{\nabla} \bm{f}
\end{equation}
With $\mathbf{x}$ as the partial coordinate, in 1D this is equivalent to
\begin{equation} \label{eq:S_equiv}
     S\bm{f} = \begin{bmatrix}
          & .\\
          & .\\
          &-c_i \frac{\partial f_i}{\partial x} \\
          & .\\
          & .\\
     \end{bmatrix}
 \end{equation}
To build the streaming matrix $C_s$, the matrix $S$ replaces $F^{(1)}$ in \cref{eq:A}, denoting these matrices as $B_m^m$. Only the diagonal transfer matrices are built, resulting in a matrix of the form: 
\begin{equation}
\label{eq:streammat}
C_s^{(k)} = \resizebox{0.35 \textwidth}{!}{
  $\begin{bmatrix}{}
  B_1^1 & 0 & 0 & 0 & 0 & \ldots & 0 & 0 & 0 & 0 \\
  0 & B_2^2 & 0 & 0 & 0 & \ldots & 0 & 0 & 0 & 0 \\
  \vdots & \vdots & \vdots & \vdots & \vdots & \ddots & \vdots & \vdots & \vdots & \vdots \\
  0 & 0 & 0 & 0 & 0 & \ldots & 0 & B_{k-2}^{k-2} & 0 & 0 \\
  0 & 0 & 0 & 0 & 0 & \ldots & 0 & 0 & B_{k-1}^{k-1} & 0 \\
  0 & 0 & 0 & 0 & 0 & \ldots & 0 & 0 & 0 & B_k^k
  \end{bmatrix}$.
}
\end{equation}
Since this process is done with the intention of a numerical simulation, second-order finite differences are used to approximate the gradient,
\begin{equation}
    c_i \frac{\partial f_i}{\partial x} \approx c_i \frac{f_i(x+1) - f_i(x-1)}{2}.
\end{equation}
In this case, the streaming matrix for one grid point has no meaning, and at least 3 grid points must be used.

\subsection{Generalization to N grid points}\label{sec:N_gridpoints}
The generalization to $N$ grid points is done separately for the collision and streaming matrices. Assume a chain of points of length $L$. All $N$-point equivalents will be denoted with math script $\mathcal{C},\mathcal{F}, \mathcal{S}, \mathcal{V} \textrm{ and } \mathcal{A}$ rather than $C,F, S, V \textrm{ and } A$. The grid point is denoted by $\mathbf{x}$ with subscript $\alpha = 1,...N$ and $[j]$ indicates the degree of the Kronecker product. Note that $\mathbf{x}$ without the subscript is the vector representing all grid points.

Introducing the new variable $\phi(\mathbf{x})$:
\begin{equation}
    \begin{split}
    \phi(\mathbf{x})&= \left(\bm{f}(\mathbf{x}_1),...,\bm{f}(\mathbf{x}_N)\right)^\mathsf{T}\\
    &=\left(f_1(\mathbf{x}_1),...,f_Q(\mathbf{x}_1),..., f_1(\mathbf{x}_N),..., f_Q(\mathbf{x}_N)\right)^\mathsf{T},
    \end{split}
\end{equation}
we construct matrices $\mathcal{S}$ and $\mathcal{F}^{(j)}$ to satisfy the following equation:
\begin{align}
\frac{\partial \phi(\mathbf{x})}{\partial t} &= \mathcal{S} \phi(\mathbf{x}) + \mathcal{F}^{(1)}(\mathbf{x}) \phi(\mathbf{x}) \notag \\
&\quad + \mathcal{F}^{(2)}(\mathbf{x}) \phi^{[2]}(\mathbf{x}) + \mathcal{F}^{(3)}(\mathbf{x}) \phi^{[3]}(\mathbf{x}),
\end{align}
that will then be used to construct the $N$-point streaming ($\mathcal{C}^{(k)}_s$) and collision ($\mathcal{C}^{(k)}_c$) matrices for:
\begin{equation}
\label{CL-ODE}
    \frac{\partial\mathcal{V}^{(k)}}{\partial t} = \mathcal{C}^{(k)}(\mathbf{x})\mathcal{V}^{(k)}(\mathbf{x}),
\end{equation}
where $\mathcal{C}^{(k)}$ is the N-point Carleman matrix defined by:
\begin{equation}
    \mathcal{C}^{(k)}=\mathcal{C}_s^{(k)} +\mathcal{C}^{(k)}_c,
\end{equation}
and $\mathcal{V}^{(k)}$ is the N-point equivalent of $V$ defined as:
\begin{equation}
\label{V-def}
    \mathcal{V}^{(k)}(\mathbf{x}) 
    =(\phi(\mathbf{x}), \phi^{[2]}(\mathbf{x}), \phi^{[3]}(\mathbf{x}), ...)^\mathsf{T}.
\end{equation}
The dimension of $\mathcal{V}^{(k)}(\mathbf{x})$ is $\sum_{j=1}^{k}N^jQ^j$ where $Q$ is the number of vector components of the lattice. $\mathcal{C}^{(k)}$ is a square matrix of dimension $(\sum_{j=1}^{k}N^jQ^j)^2$.

The explicit construction for the $N$ grid point collision and streaming matrices can be found in \cref{sec:clin_appendix}.

\section{Boundary and Initial conditions} \label{sec:bcic}

\subsection{Boundary conditions}\label{sec:Boundary_Conditions}
In our analysis, we chose periodic boundary conditions for simplicity. They are applied in space by setting:
\begin{align}
    g (\mathbf{x}_{N+1}) = g(\mathbf{x}_1)
\end{align}
where $N$ is the number of grid points. This condition holds for any equation $g$. This condition was imposed on the equations governing the streaming and collision operators, which were then translated into the exact entries of the respective matrices. The boundary conditions therefore appear in the Carleman matrix. 

For the collision operator, when expanding for $N$ grid points, the transformation at each grid point is not affected by any others, since the transformation in \cref{sec:Ncollmat} is:
\begin{align}
    \mathcal{F}^{(j)}(\mathbf{x}_{\alpha}) = \delta_{\alpha}^{[j]} \otimes F^{(j)},
\end{align}
 and so $\mathcal{F}(\mathbf{x}_{N+1})$ doesn't need to be taken into account as it does not affect $\mathcal{F}(\mathbf{x}_{1}), \mathcal{F}(\mathbf{x}_{2}), ....\mathcal{F}(\mathbf{x}_{N})$. Hence, the $N$-point collision matrix $\mathcal{C}_c^{(3)}$ requires no modification.

For the streaming matrix, as described in \cref{sec:Nstreaming},
the gradient operator is approximated by second-order finite differences which naturally lends itself to imposing any set of boundary conditions, including periodic. 
For the construction of a streaming matrix with periodic boundary conditions refer to \cref{sec:Periodic_BC}.

\subsection{Initial conditions} \label{sec:initial_conditions}
The initial conditions are applied after the Carleman linearization and the generalization to $N$ grid points.
To impose some initial configuration on the system, i.e. to define the initial heights and velocities at each grid point, we consider:
\begin{equation}
    \mathcal{V}^{(3)}\left(t_0, \mathbf{x}\right)=(\phi(t_0,\mathbf{x}), \phi^{[2]}(t_0,\mathbf{x}), \phi^{[3]}(t_0,\mathbf{x}))^\mathsf{T},
\end{equation}
where
\begin{align}
    \phi(t_0,\mathbf{x})=\left(\bm{f}(t_0,\mathbf{x}_1),...,\bm{f}(t_0,\mathbf{x}_N)\right),
\end{align}
and
\begin{align}
    \begin{split}
    \bm{f}(t_0, \mathbf{x}_{\alpha}) &= \bm{f}^{\rm eq}(\mathbf{x}_{\alpha}) \\ &=h_{\alpha}\left(\frac{2}{3}, \frac{1}{6}+\frac{u_{\alpha}}{2}, \frac{1}{6}-\frac{u_{\alpha}}{2}\right),
    \end{split}
\end{align}
which accounts for our choice to model a D1Q3 system. In the equation, ${h_{\alpha}}$ is the normalized initial height distribution at grid point ${\alpha}$ and $u_{\alpha}$ is the initial velocity at grid point ${\alpha}$.  

\section{Forward Euler}\label{sec:forward_euler}

The next step is to transform the ODE into a linear system of equations by using an appropriate discrete approximation for the time derivative in \cref{eq:DVBE}. This is done using the forward Euler approximation, which is essentially a first-order expansion in time.

In this section, we will apply the high-order method \cite{berry2017quantum} and the truncation that realizes the Forward Euler approximation \cite{berry2014high} to the matrix $\mathcal{C}^{(3)} \equiv \mathcal{C}$ and vector of Carleman variables $\mathcal{V}^{(3)} \equiv \mathcal{V}$ introduced in \cref{sec:clin}.

We begin with an initial value problem consisting of the result of the collision-streaming step obtained by the Carleman linearization, which is a first-order differential equation based on some initial configuration:

\begin{align}
\left\{
\begin{aligned}
\mathcal{V}_0 & = \mathcal{V}(t=0) \\
\frac{d \mathcal{V}}{d t} &= \mathcal{C}\mathcal{V}.
\end{aligned}
\right.
\end{align}

Since $\mathcal{C}$ is time-independent, the exact solution of the initial value problem is given by:
\begin{align}
  \mathbf{\mathcal{V}}(t) = e^{\mathcal{C}t} \mathbf{\mathcal{V}}_0. 
\end{align}

Applying a Taylor expansion of order k to this result then yields:
\begin{align}
  e^{\mathcal{C}t} \approx \sum_{j=0}^{k} \frac{(\mathcal{C}t)^j}{j!} \equiv  T_k(\mathcal{C}t). 
\end{align}
 
By henceforth assuming the evolution timestep to be small compared to the flow characteristic time $\mathcal{T}$, i.e. $\delta t/\mathcal{T} \ll 1$, and using a large order expansion \(k\), we can approximate the solution of the evolved configuration as:
\begin{align}
  \mathbf{\mathcal{V}}(t) \approx T_k(\mathcal{C}\delta t) \mathbf{\mathcal{V}}_0.  
\end{align}
This solution is then iteratively evolved many times to reach the total evolution time. Every configuration evolved for the timestep  \(\delta t\) becomes the initial configuration for the next evolution. This procedure is repeated \(T/\delta t\) times, where $T$ is the total evolution time.
\begin{align}
\mathbf{\mathcal{V}}_{j+1}  = T_k(\mathcal{C}\delta t) \mathbf{\mathcal{V}}_j,
\end{align} 
where $\mathcal{V}_j$ is used to denote $\mathcal{V}(j \cdot \delta t)$. Choosing $k=1$, we impose a first-order approximation, this is the key step to realize the Forward Euler Approximation. So the equation becomes 
\begin{align}
\mathbf{\mathcal{V}}_{j+1} - \left(I + \mathcal{C}\delta t\right) \mathbf{\mathcal{V}}_j = 0.
\end{align}
Following the reasoning of evolving small timesteps with a low-order expansion, we can construct a matrix of the following form:
\begin{align}
\resizebox{0.42 \textwidth}{!}{\ensuremath{
E=
\begin{bmatrix}
I & 0 & 0 & 0 & \cdots & 0 \\
-\left(I + \mathcal{C}\delta t\right) & I & 0 & 0 & \cdots & 0 \\
0 & -\left(I + \mathcal{C}\delta t\right) & I & 0 & \cdots & 0 \\
0 & 0 & -\left(I + \mathcal{C}\delta t\right) & I & \cdots & 0 \\
\vdots & \vdots & \vdots & \vdots & \ddots & \vdots \\
0 & 0 & 0 & 0 & \cdots & I
\end{bmatrix}}.
}
\end{align}

The matrix $E$ can then be used to create an approximate LSE for the ODE given in \cref{CL-ODE} as follows:

\begin{align}\label{LSE}
E
\begin{bmatrix}
\mathcal{V}_0 \\
\mathcal{V}_1 \\
\mathcal{V}_2 \\
\vdots \\
\mathcal{V}_n
\end{bmatrix}
=
\begin{bmatrix}
\mathcal{V}_0 \\
0 \\
0 \\
\vdots \\
0
\end{bmatrix},
\end{align}
where $\mathcal{V}_j$ denotes the N-grid point distribution function vector as defined in \cref{V-def} after $j$ timesteps and $\mathcal{V}_0$ the initial configuration.

The Euler method yields an error that scales as $\mathcal{O}(\delta t^2)$ for a single timestep. Therefore, the error in the total simulation amounts to $\mathcal{O}(N_t \delta t^2) = \mathcal{O}(\Delta \delta t^2 / N_t) \approx  \mathcal{O}(\delta t)$ . To achieve an error bounded by $\epsilon$, we impose $N_t = \mathcal{O}(\Delta \delta t^2 / \epsilon)$ \cite{berry2014high}.

The matrix is square and its two dimensions scale as \((3n + 9n^2 + 27n^3)(N_t + 1)\), with $n$ number of grid points and $N_t$ number of timesteps, i.e. as \( \mathcal{O} (n^3 N_t)\), as described in \cref{sec:N_gridpoints}. This stems from the fact that the Carlemann matrix is repeated \(N_t\) times on the second block diagonal.
The matrix is sparse, with the sparsity pattern shown in \cref{fig:sparsity_pattern} for the case of \(n = 3\) and \(N_t = 4\):
\begin{figure}[htbp]
    \centering
    \includegraphics[width=0.9\columnwidth]{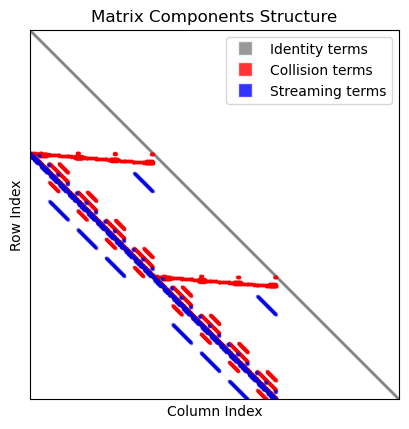} 
    \caption{Sparsity pattern of the matrix after Forward Euler Approximation with $Timesteps = 3$. The collision terms are shown in red and the streaming terms are shown in blue.}
    \label{fig:sparsity_pattern}
\end{figure}
The physical observables, such as height and streaming velocity, can be extracted at each timestep by summing over the distribution functions $f_i$ as shown in \cref{eq:mom_0} and \cref{eq:mom_1}.

The result vector contains many stacked subvectors \(x_0\), \(x_1\),... from which we can extract three elements in the vector every \(3n + 9n^2 + 27n^3\) elements, with $n$ number of grid points, to obtain the distribution function \(f_1\),\(f_2\), \(f_3\), which then allows us to calculate the heights and velocities. All other values in the solution vector are required to capture the nonlinear dynamics for the next timestep (as approximated via the Carleman linearization), but do not directly relate to the physical observables of the system. Thus, we only need a small fraction of the elements that compose the full vector.

\section{Classical Benchmarking}\label{sec:classical_benchmarking}

To benchmark our linearized system of equations, we employed the ETH Euler cluster to compute solutions and extract relevant parameters.

In this section, the system of equations is solved classically with the numpy function $\texttt{np.linalg.inv()}$. Then the inverted matrix is multiplied with the vector that encodes the initial configuration with $\texttt{np.dot()}$.

To validate the correctness of our approach, we tested it using different initial configurations. The limited number of grid points and timesteps used in these tests reflects both the constraints of available computational resources and the exponential growth of the system's dimensionality with increasing resolution.

The number of grid points directly determines the resolution with which the system's dynamics can be modeled. As such, meaningful results are only expected for test cases where the dynamics scale appropriately with the resolution of the discretization. Before assessing the accuracy of our model by comparing its outputs to known analytical solutions, it is thus essential to quantify the limitations imposed by the relatively coarse discretization used in our tests.

\subsection{Stable configuration test}\label{sec:configuration_test}

To evaluate the expected error for a given test case and spatial extent, we first performed a benchmark test using a stable steady-state configuration. In this setup, the water height is set to be constant across the entire domain at the initial timestep. As the system evolves over time, numerical errors arising from the limited resolution gradually accumulate, leading to deviations from the steady-state configuration.

The error was calculated as the sum of the deviations at each grid point, normalized by the total number of grid points, yielding a measure of relative error. This metric allows us to estimate the stability and accuracy of the scheme for different system sizes.

\begin{figure}[tbp]
    \centering
    \includegraphics[width=0.9\columnwidth]{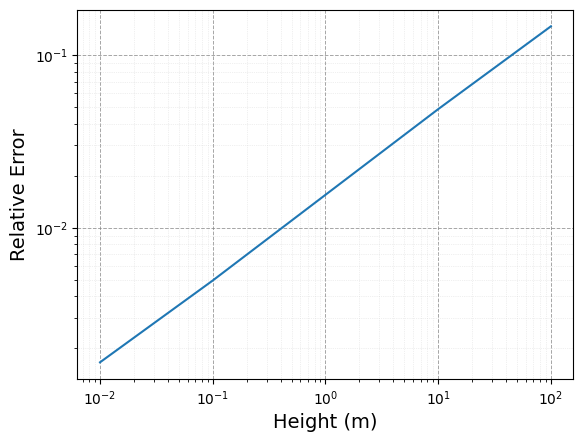}
    \caption{Relative error vs. initial height for a configuration where the height is constant across the entire domain. The error is shown after $N_t=4$ timesteps using 4 grid points.}
    \label{fig:Rel_err_vs_height}
\end{figure}

In the graph, we observe a linear increase in the relative error. As expected, the scheme maintains low relative errors for smaller water heights. Specifically, the simulation shows sufficient accuracy within 0.01 and 0.1~m. Configurations in this range are therefore good candidates to compare to analytical solutions. 

\subsection{Normal models propagation speeds: Speed of sound}\label{sec:speed_of_sound}

To further verify the accuracy of our approach within the identified stable range, we conducted a second test involving the evolution of a pressure/density step perturbation $\delta h_0$. This configuration can be used to measure the speed of sound in the system effectively validating the pressure and showing that the model captures the proper dispersion of eigen-modes in the hydrodynamic limit in a linear regime. To that end, the perturbation magnitude should remain small, i.e. $\delta h_0/h_0<<1$. In the linear regime, the analytical solution of the SWE for such a configuration predicts a propagation speed of \( v = \sqrt{gh_0} \), where \( h_0 \) is the initial water height and \( g \) is the gravitational acceleration constant~\cite{bhl58383}.

In the initial setup, the domain was divided evenly, with half of the grid points set to a height \( h_0 \) and the other half to \( h_0 + \delta h_0 \) with $\delta h_0 = 0.01 h_0$. This configuration created a step function wave suitable for evaluating the scheme’s ability to capture wave propagation dynamics accurately.  
To calculate the velocity of the propagating wave, we measured the number of timesteps \( n_t \) it took for the wave peak to travel to the next grid point. The velocity was then computed using the formula:  
\begin{align}
v = \frac{L / N_{\text{grid}}}{n_t \delta t},
\end{align}  
where \( L \) is the total length of the domain, \( N_{\text{grid}} \) is the number of grid points, and \( \delta t \) is the size of each timestep, which we chose as a function of the height \( h_0 \).  

\begin{figure}[tbp]  
    \centering  
    \includegraphics[width=0.9\columnwidth]{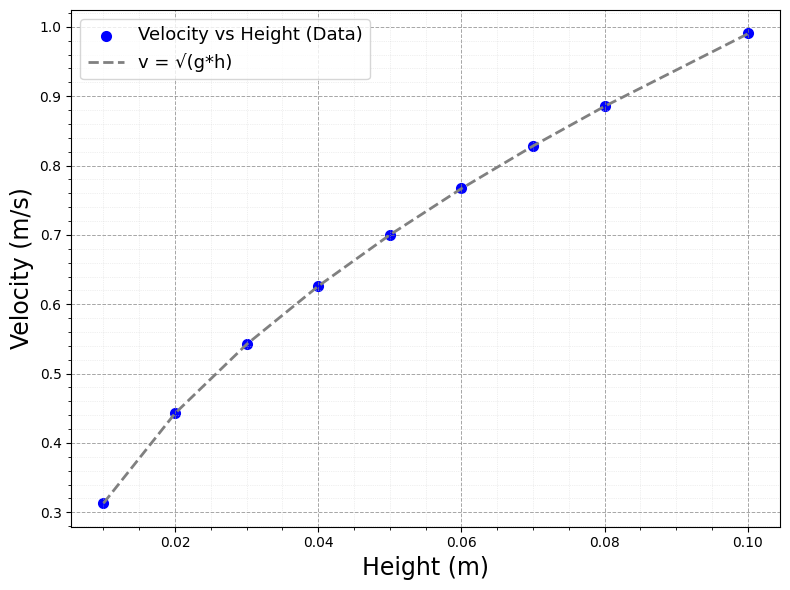} 
    \caption{Velocity of the propagating wave vs. initial height.}
    \label{fig:velocity_vs_height}
\end{figure}  

The measured velocity clearly exhibited the expected square-root scaling, confirming that our numerical scheme successfully captures the fundamental dynamics of the intended fluid simulation. Although our resolution limitations prevented us from modeling more complex evolutions, the fact that our mapping can reliably demonstrate key features of the SWE provides good evidence for its conceptual validity. The code that realizes the mapping, used to obtain these plots, can be found on GitHub~\cite{quantech-swe-qlss-code}.

\section{Quantum linear system solvers}\label{sec:qlss}

The linear system obtained by the linearization scheme described in the previous sections can be solved efficiently on a quantum computer under certain conditions. The first quantum algorithm for solving linear systems, introduced by Harrow, Hassidim, and Lloyd (HHL)~\cite{harrow2009quantum}, demonstrated that quantum computers could, in principle, solve linear systems of equations with complexity logarithmic in the system dimension, potentially providing an exponential speedup over the best-known classical methods.

\subsection{Quantum input–output model}\label{sec:quantum_io_model}

In contrast to the classical setting where the output of an algorithm is an explicit numerical vector, the quantum input–output model represents both the vector $b$ and solution vector $x$ as quantum states $|b\rangle$ and $|x\rangle$ in a Hilbert space $\mathcal{H} = (\mathbb{C}^2)^{\otimes n}$, which is the state space of an $n$-qubit quantum computer. The objective is to prepare a quantum state  
\[
	\ket{x} \propto A^{-1}\ket{b}
\]
for a given linear system $A x = b$.  The error in the solution is quantified by  
\[
	\epsilon = \| \ket{x} - \ket{\tilde{x}} \|,
\]
where $|\tilde{x}\rangle$ represents the algorithm's output state and $\| \cdot \|$ denotes the Hilbert space norm induced by the inner product on $\mathcal{H}$.

Quantum algorithms manipulate these states through unitary transformations, which can be realized efficiently using quantum gates~\cite{nielsen2010quantum}. While quantum computers natively implement unitary evolutions, we often have to manipulate matrices, such as $A$, which are generally nonunitary. To this end, the technique of block-encoding embeds $A$ as the top-left block of a larger unitary operator $U_A$:
\[
	U_A = \begin{pmatrix} A/\alpha & \cdot \\ \cdot & \cdot \end{pmatrix}
\]
where $\alpha$ is a normalization factor ensuring the singular values of $A$ are bounded by 1. This technique allows quantum operations on $U_A$ to induce the desired transformations on $A$ itself.

\subsection{Quantum singular value transformation}\label{sec:qsvt}

Since HHL, newer and more efficient algorithms have been developed that achieve (near-)optimal dependencies on parameters such as the condition number $kappa$ and the target precision $\epsilon$, as summarized in \cref{tab:qlss_summary}. The condition number $\kappa$ is defined as the ratio between the largest and the smallest singular values of $A$: $\kappa := \sigma_{max}/\sigma_{min}.$ It determines the eigenvalue range of the (rescaled) block-encoded matrix $A$. 

Our system can be solved using any of these near-optimal methods.  Here, we focus on the method based on the quantum singular value transformation.  While its runtime dependence on the condition number $\kappa$ is suboptimal compared to the very latest algorithms, QSVT clearly illuminates the central role of the condition number in the algorithm's performance~\cite{lin2022lecturenotesquantumalgorithms, Martyn_2021}. 

\begin{table*}[t]
    \centering
    \begin{tabular}{|l|l|l|l|}
        \hline
        \textbf{Algorithms} & \textbf{Characteristics} & \textbf{Runtime Complexity} \\
        \hline
        HHL \cite{harrow2009quantum} & First QLSS algorithm & $O(\log(N) s^2 \kappa^2/\epsilon)$ \\
        \hline
        QSVT \cite{gilyen2019quantum} & Block-encoding framework & $O(\kappa^2 \log(\kappa/\epsilon))$ \\
        \hline
        Discrete adiabatic \cite{costa2022optimal} & Optimal scaling of $\kappa, \epsilon$ & $O(s\kappa \log(1/\epsilon))$ \\
        \hline
        Augmentation and kernel reflection \cite{dalzell2024shortcut} & No trial state dependency & $O(s\kappa \log(1/\epsilon))$ \\
        \hline
    \end{tabular}
    \caption{Comparison of key quantum linear system solvers. Runtime complexity is expressed in terms of condition number $\kappa$, system size $N$, sparsity $s$, and precision $\epsilon$. Table adopted from a recent review \cite{morales2024quantum}.}
    \label{tab:qlss_summary}
\end{table*}

QSVT is a meta-algorithm for transforming singular values of a block-encoded matrix through polynomial functions. A matrix function $f(A)$ for a diagonalizable matrix $A$ with eigendecomposition $A = S\Lambda S^{-1}$ is defined as $f(A) = Sf(\Lambda)S^{-1}$, where $f(\Lambda)$ applies the scalar function $f$ to each eigenvalue. For solving linear systems, the inverse function $f(A) = A^{-1}$ is of particular interest because solving $A x = b$ requires applying the inverse $A^{-1}\ket{b}$. However, QSVT can only implement polynomial functions of matrices. Therefore, we must first construct a polynomial approximation $p(\lambda) \approx 1/\lambda$ over the spectrum of $A$.

The QSVT-based solver approach involves several conceptual stages:

\begin{itemize}
    \item \textit{Encoding the matrix}: Instead of directly working with $A$, we embed it within a unitary matrix, as required for quantum computing.
    \item \textit{Singular value transformation:} Apply QSVT to transform singular values \( \sigma_i \) of $A$ using a polynomial approximation of \( f(x) = 1/x \).
    \item \textit{Quantum state preparation:} Prepare the quantum state \( |b\rangle \) corresponding to the right-hand side of the equation.
    \item \textit{Application of QSVT:} Apply the QSVT-derived unitary transformation to obtain \( |x\rangle \), the quantum state representing the solution.
    \item \textit{Measurement and post-processing:} Measure and extract useful information from \( |x\rangle \), often requiring classical post-processing.
\end{itemize}

\subsection{Numerical results} \label{sec:quantum_numerics}

\begin{figure}[bp]
    \centering
    \includegraphics[width=0.9\columnwidth]{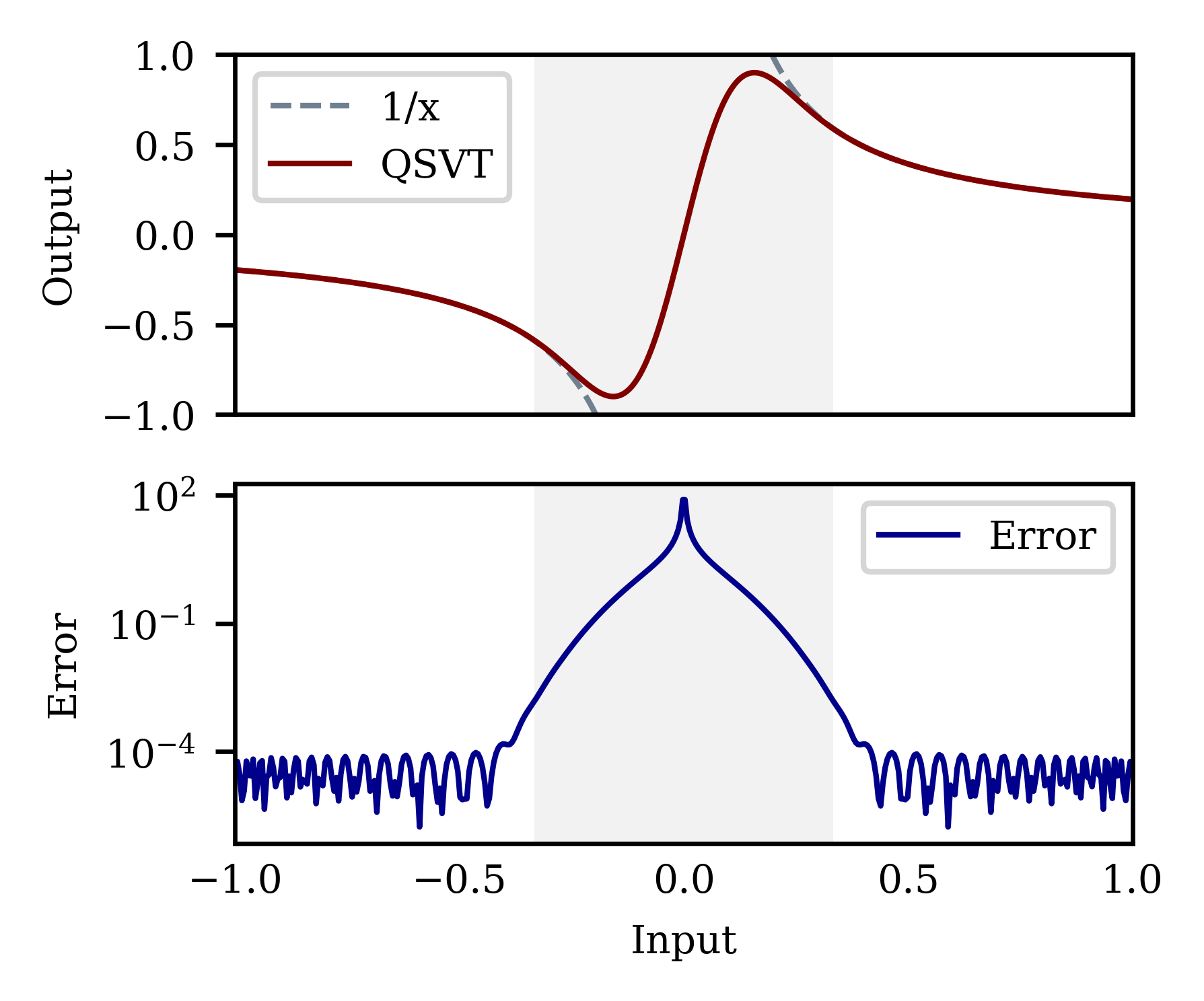}
    \caption{QSVT approximation of the inverse function. The top panel shows accurate approximation (red) in the white region, with errors spiking in the irrelevant gray region. The bottom panel quantifies the error (blue). Numerical analysis was performed using the \texttt{pyqsp} package~\cite{Martyn_2021}.}
    \label{fig:qsvt_plot}
\end{figure}

The key parameter defining the complexity and runtime of the linear system solver using QSVT is the condition number $\kappa$. In the top panel of \cref{fig:qsvt_plot}, the red function approximates the inverse function accurately within the white region, while errors spike in the gray region, as shown in the bottom panel. However, the gray region is irrelevant since the rescaled eigenvalues lie outside of it. A larger $\kappa$ means the inverse function must be approximated accurately in the broader domain, necessitating a higher-degree polynomial, and ultimately, a deeper quantum circuit~\cite{Martyn_2021}. 

\begin{figure*}[!tbh]
    \centering
    \includegraphics[width=1.9\columnwidth]{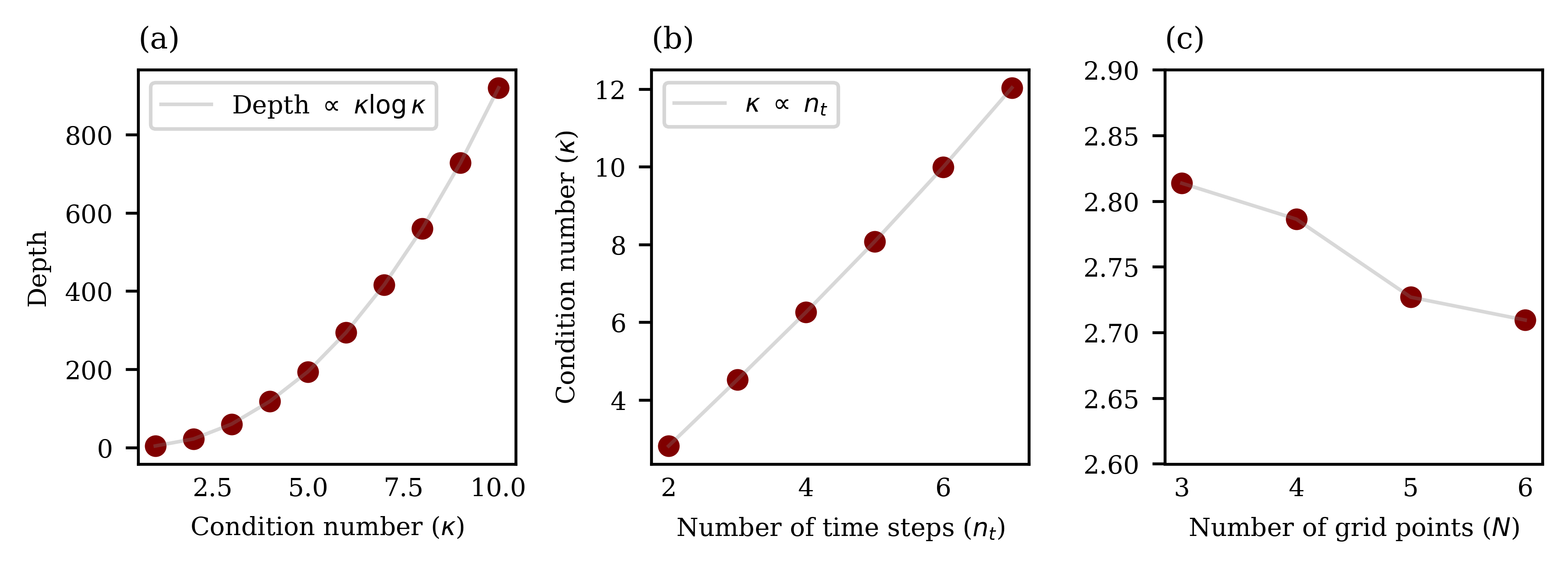}
    \caption{Scaling of the condition number $\kappa$ in QSVT. (a) Required polynomial degree scales as $\mathcal{O}(\kappa\log\kappa)$. (b) $\kappa$ grows linearly with time steps $n_t$. (c) $\kappa$ remains nearly constant with grid points $N$. Numerical analysis was performed using the \texttt{pyqsp} package~\cite{Martyn_2021}}
    \label{fig:qsvt_scaling}
\end{figure*}

More precisely, the required polynomial degree scales as $\mathcal{O}(\kappa\log(\kappa))$~\cite{lin2022lecturenotesquantumalgorithms}, as we numerically verified in \cref{fig:qsvt_scaling} (a). Furthermore, we verify that $\kappa$ scales linearly with the number of time steps, as shown in \cref{fig:qsvt_scaling} (b). The linear scaling with the number of time steps aligns with the theoretical prediction~\cite{berry2017quantum}.
Finally, to analyze how $\kappa$ grows with the number of grid points, we performed numerical simulations for $3, 4, 5,$ and $6$ grid points. Interestingly, \cref{fig:qsvt_scaling} (c) reveals a non-trivial behavior where $\kappa$ appears almost independent of the number of grid points. A priori, it is unclear how $\kappa$ should scale with grid resolution. This non-trivial behavior currently lacks a clear explanation and warrants further investigation. The code implementing the QSVT algorithm used to obtain these plots can be found on GitHub~\cite{quantech-swe-qlss-code}.

In summary, our numerical analysis did not reveal any bottlenecks that would negate a potential exponential speedup, at least within the quantum part of the computational pipeline. This suggests that the approach remains theoretically sound and computationally feasible.

\subsection{Input and readout}\label{sec:input_readout}
A practical end-to-end application of QSVT for solving a linear system depends on two critical assumptions. First, an efficient block encoding of the classical matrix $A$ and efficient state-preparation of the initial state $\ket{b}$ are required. Second, meaningful information must be efficiently extracted from an exponentially large quantum state~\cite{aaronson2015read, dalzell2023quantum}.

\subsubsection{Readout problem}

A complete description of a quantum state requires full state tomography, which is exponentially hard. However, in cases where one is only interested in a restricted region of space — such as a fixed number of grid points necessary to describe a localized phenomenon, like a wave collision with an object -- restricted state tomography may be performed efficiently.  

Alternatively, one can estimate a global property of the state, such as a particular coefficient of interest, by measuring a suitable observable. These approaches allow for extracting meaningful results without negating the potential speedup.

\subsubsection{Encoding classical data}

The second crucial assumption concerns encoding classical data, which consists of two main tasks. First, the initial quantum state $\ket{b}$ must be prepared efficiently. In general, preparing an arbitrary $n$-qubit state forces either the circuit depth or the number of ancilla qubits to grow exponentially in $n$~\cite{sun2023asymptotically, zhang2022quantum}. However, for some initial states that exhibit additional structure, the exponential cost can be avoided. For example, when the amplitudes are defined by an efficiently computable function -- such as a probability density (e.g., a Gaussian)~\cite{grover2002creating, kitaev2008wavefunction, rattew2021efficient} or a real-valued signal approximated by a low-degree polynomial~\cite{gonzalez2024efficient, iaconis2024quantum} -- there exist efficient state preparation algorithms. Likewise, if the state is $k$-sparse -- that is, it has only $k$ nonzero amplitudes -- one can build a circuit using $\mathcal{O}(\log{nk})$ circuit depth and $\mathcal{O}(nk\log{k})$ auxiliary qubits~\cite{zhang2022quantum}. Therefore, there exist structured settings, including physically relevant applications, where the initial state $\ket{b}$ can be loaded efficiently.

The more significant challenge is obtaining a block encoding of the matrix $A$. This typically requires quantum random access memory (QRAM), whose initialization is exponentially hard in both gate and qubit complexity~\cite{jaques2023qram}. However, two factors mitigate this challenge. Firstly, optimizations exist for sparse matrices, making QRAM more practical~\cite{lin2022lecturenotesquantumalgorithms}, and, secondly, QRAM initialization needs to be performed only once. Once QRAM is set up and polynomial query access to the block encoding is available, the initial value problem can be solved efficiently for multiple initial states. In contrast, most classical methods require solving the problem from scratch for each new initial condition, making the quantum approach potentially more efficient.

\section{Conclusion}\label{sec:conclusion}
Our work explored the application of quantum algorithms to fluid dynamics by adapting and benchmarking a linearization scheme for the shallow water equations previously developed for the Navier-Stokes equations. Unlike the NSE, which describes an evolution in pressure and velocity and requires the assumption of weak compressibility~\cite{li2025potential}, when applying the mapping to the SWE, we had to assume small variations in water elevations relative to the mean fluid depth, as for example found in surface dynamics in open-ocean systems. Furthermore, our classical simulations confirmed the method’s general validity and highlighted its potential, although due to the large computational cost of this scheme for classical computers, open questions remain in terms of stability and behavior for larger system sizes. 

Another key open challenge remains in the efficient extraction of information from quantum states. Since full state tomography is exponentially hard, reading out all physical quantities remains inefficient. However, for cases where only a fraction of the information is of interest, promising alternatives include restricted state tomography over a small set of grid points and quantum expectation estimation to extract physical observables such as mean energy. In addition, the problem of encoding classical data into quantum states warrants further investigation. Efficient state preparation schemes are needed to avoid the exponential overhead associated with unstructured state initialization. Despite these challenges, our results provide a foundation for further research into quantum-enhanced computational fluid dynamics, particularly as quantum hardware continues to advance.

\paragraph*{Acknowledgments:} We are grateful to Prof. Ilya~Karlin for insightful discussions on the lattice-Boltzmann method and computational fluid dynamics, and for his valuable feedback on the manuscript. We also thank Dr. Gabriele Raino for bringing together the authors by organizing the Quantech workshop at ETH Zurich and for supporting this project’s development. Portions of this manuscript were drafted or edited with the assistance of ChatGPT to improve clarity and style.

\appendix
\section{Construction of Carleman matrices for N grid points} \label{sec:clin_appendix}

\subsection{Collision matrix (N grid points)}\label{sec:Ncollmat}
For the collision matrix, first define for each tensor degree and each grid point:
\begin{equation}
    \mathcal{F}^{(j)}(\mathbf{x}_{\alpha}) =\delta_{\alpha}^{[j]}\otimes F^{(i)}
\end{equation}
with 
\begin{equation}
    \delta_{\alpha} = \underbrace{(0,...,\overbrace{1}^{\alpha- \textrm{th position}}...,0)}_{N \textrm{ elements}}.
\end{equation}
Then for the degree $(j)$, we define a vector for all grid points:
\begin{equation}
    \mathcal{F}^{(j)}(\mathbf{x}) = \left(\mathcal{F}^{(j)}(\mathbf{x}_{1}),..., \mathcal{F}^{(j)}(\mathbf{x}_{N})\right)^\mathsf{T}
\end{equation}
We further define a replacement transfer matrix for \cref{eq:A}, where $\mathcal{F}^{(i)}(\mathbf{x})$ replaces $F^{(j)}$:
\begin{equation}
\label{eq:NA}
\resizebox{0.9\columnwidth}{!}{
$\mathcal{A}_{j+m-1}^i(\mathbf{x}) = 
  \sum_{r=1}^j 
  \overbrace{
    \mathbb{I}_{NQ\times NQ} \otimes \cdots \otimes 
    \underset{\substack{\uparrow \\ \text{r-th position}}}{\mathcal{F}^{(m)}(\mathbf{x})} 
    \otimes \cdots \otimes \mathbb{I}_{NQ\times NQ}
  }^{j\text{ factors}}$
}
\end{equation}

This definition of $\mathcal{A}$ replaces $A$ in \cref{eq:collmat} which then gives the N-point collision matrix $\mathcal{C}^{(k)}_c$.
\subsection{Streaming matrix (N grid points)}\label{sec:Nstreaming}
To create the N-point streaming matrix we need to construct $\mathcal{S}$. This matrix satisfies the equation:
\begin{equation}
    \mathcal{S} \phi(\mathbf{x}) =(...,-c_i \partial f_i(\mathbf{x}_{\alpha})/\partial_x, ...)^\mathsf{T}
\end{equation}
where we use second-order finite differences to approximate the spatial gradient:
\begin{equation}\label{gradient}
    \frac{\partial f_i(\mathbf{x}_{\alpha})}{\partial x} = \frac{(N-1)\left(f_i(\mathbf{x}_{\alpha +1}) - f_i(\mathbf{x}_{\alpha -1})\right)}{2L}
\end{equation}
As we limit our analysis to the 1D case, we only need to calculate the gradient for $\partial x$. Also note that there must be a minimum of 3 grid points to use second order finite differences. Boundary conditions are treated in the next chapter.

The matrix $\mathcal{S}$ replaces $\mathcal{F}^{(1)}$ in \cref{eq:NA}, thus giving transfer matrices denoted as $\mathcal{B}^j_j$ that sit on the diagonal of $\mathcal{C}_s^{(k)}$, while neglecting transfer matrices on the off-diagonals. This gives the N-point streaming matrix $\mathcal{C}_s^{(k)}$.

\subsection{Boundary condition matrix}\label{sec:Periodic_BC}
The resulting matrix has 3 diagonals with non-zero elements and is as follows:
\begin{align}
\mathcal{C}_s^{(3)}=\frac{-1}{\delta t}\cdot
\begin{bmatrix}
0 & 0 & 0 & 0 & 0 & \cdots & 0 & 0 & 0 \\
0 & 1 & 0 & 0 & -1 & \cdots  & 0 & 1 & 0 \\
0 & 0 & -1 & 0 & 0 & \cdots & 0 & 0 & 1 \\
0 & 0 & 0 & 0 & 0 & \cdots  & 0 & 0 & 0 \\
 &  &  &  &   \ddots & & &&\\
0 & 0 & 0 & 0 & 0 & \cdots  & 0 & 0 & 0 \\
0 & -1 & 0 & 0 & 0 & \cdots  & 0 & 1 & 0 \\
0 & 0 & 1 & 0 & 0 & \cdots  & 0 & 0 & -1 \\
\end{bmatrix}
\end{align}
Here the periodic boundary condition are implemented in the first and in the last 3 rows, where the elements belonging to the third diagonals are copied on the other side of the matrix. The coefficient $-\frac{1}{\delta t}$ comes from the definition of the vector of velocities $c_i$ and the gradient.

\newpage
\bibliography{bibliography}

\end{document}